# PROBING EXOPLANETARY MAGNETISM VIA ATOMIC ALIGNMENT EFFECT

M. Rumenskikh, A.V. Taichenachev, I.F. Shaikhislamov, V.I. Yudin

**Abstract**

The intrinsic magnetic fields of exoplanets affect the structure of their atmospheres and plasmaspheres and, therefore, the observational manifestations of transit absorptions. This work proposes a new method for constraining the presence or absence of relatively weak magnetic fields. The method is based on the quantum effect of atomic alignment of the lower energy level resulting in changing the absorption probabilities of individual transitions of multiplets from the equilibrium $2J+1$ value. It appears to be sensitive to fields above $\sim 0.001$ G. We applied this method to some available transit observations of exoplanets and demonstrate that we indeed have the possibility to constrain the intrinsic magnetic field of some exoplanets right now. However, more precise and repetitive measurements, which might be available in near future, are needed for definite conclusions.

**Introduction**

The development of exoplanetary research driven by the launch of new ground and space missions, is moving forward at a breathless pace. Already more than 5000 exoplanets have been discovered. Transit absorptions confirmed the presence of atmospheres and this allows more detailed study of planetary systems. Observations of transit absorptions have shown that many exoplanetary atmospheres absorb radiation from the parent star at wavelengths corresponding to the lines of hydrogen, helium and heavy elements. Detailed analysis based on numerical modeling (Bourrier +, 2014; Oklopcic +, 2018; Khodachenko +, 2021; Odert +, 2020; Lampon +, 2020; Ben - Jaffel +, 2022) showed that the parameters of radiation and the plasma wind of the parent star largely determine the shape of the spectral lines formed in upper atmospheres of exoplanets.

The composition of the atmosphere and its structure vary significantly depending on many factors such as spectral class of the parent star, age of the system, planetary orbit, mass and radius, etc. The remarkable diversity of exoplanets and the breadth of observational data available have provided an unparalleled opportunity to study the architecture and evolution of planetary systems. Accurate characterization of exoplanetary atmospheres, however, hinges on detailed analysis of optical data obtained from observations by telescopes.

Besides other factors, the absence of information about a planet's magnetic field (MF) poses a significant challenge for the accurate modeling of atmospheric properties of exoplanets. Magnetic fields significantly change or even determine the dynamics of partially ionized atmosphere and evolution, influencing observational signatures. Planetary magnetic field might be crucial for potential habitability, shielding planetary surfaces from harmful stellar radiation. Furthermore, knowledge of exoplanets with constrained intrinsic magnetic field will facilitate the progress in theories of planetary magnetism.

The determination of magnetic field of exoplanets is a critical in the study of planetary systems for several reasons:

- Impact on atmospheric structure. MFs significantly influence the structure and dynamics of plasmaspheres, directly affecting the interpretation of transit absorption spectra.

- Support for Dynamo theory. Expanding the empirical basis for planetary magnetic field observations offers opportunities to refine and validate dynamo theory.
- Potential biomarker. MFs may serve as an indirect biomarker, as they are instrumental in shielding planetary surfaces from harmful stellar radiation, thus playing a role in habitability.

To date, the only viable direct method of detecting magnetic field of exoplanets rely on the radio-emission of charged particles within their atmospheres subject to Larmor gyration. A number of studies have explored this mechanism and its observational manifestations. Prominent work by Zarka (2007) and Lynch+ 2018 offered optimistic predictions for radio emission from hot Jupiters, estimating radiation levels that could be detected against stellar background, assuming the probable value of intrinsic magnetic field. Despite these predictions, observational campaigns such as those conducted with the LoFAR radio telescope have yielded negative results so far. For instance, only faint inconclusive radio emissions were detected from the planetary system tau Bootis b, as reported by Turner+ (2021, 2024), while other targets exhibited no significant signatures.

An alternative method, presented in this paper, was inspired by earlier work of Varshalovich (1970), who described the determination of magnetic field strength and direction in comets via sodium doublet absorption in their tails. The effect is based on atomic alignment and arises from the interaction of absorbing atoms with both magnetic fields and solar radiation. This interaction alters the population of magnetic substates of levels and, consequently, the relative depth of absorption in doublet lines. This idea was later applied by Yan & Lazarian (2008) to the weak magnetic fields in diffuse astrophysical environments.

Exoplanetary upper atmospheres, with their relatively low densities, offer a promising environment for applying atomic alignment theory. Deviations in the absorption strengths of individual lines in multiplets from their equilibrium ratios $(2J + 1)$ can be caused by the alignment effect, and thus, also subject to magnetic field. From the start it is necessary to clarify that the effect considered in this paper is not related directly to the well known Zeeman splitting of spectral lines. The Zeeman effect also has the observational manifestations, but of a completely different nature. Quantitatively, for a reference field of ~1 G the line splitting is on average 5 orders of magnitude smaller than the typical width of the spectral lines that we observe at exoplanet atmospheres. Thus, it is unlikely to be distinguished in the near future. In the upper atmospheres of hot exoplanets with typical temperature of $10^4$ K and flow velocity of ~10 km/s the broadening of absorption lines occurs mainly due to the Doppler effect.

The method proposed in this study is based on the quantum phenomenon of atomic alignment – change of angular momentum projections of electron at a given quantum level due to anisotropic radiation of the host star. The anisotropy is a natural consequence of radiation being directed from a star to a planet. This alignment effect requires some specific conditions detailed in the Model Description section. In the absence of a planetary magnetic field (MF), the absorption of directed flux of stellar photons initiates transitions between sublevels with different probabilities (Fig.1) that modifies the population ratios of sublevels of the bottom level in accordance with the atomic alignment effect. This leads to deviation of absorption seen by observer in atmospheric multiplet lines from the equilibrium $2J+1$ ratio. However, the inclusion of an MF disrupts this alignment by rotating the quantization axis and reducing anisotropy. Above a certain MF intensity and at orientations not parallel to radiation, the absorption ratio in multiplet lines reverts to the equilibrium $2J+1$ values. In the simplest form, this approach, therefore, can be applied for identifying the absence of intrinsic planetary magnetic fields, by measuring with sufficient accuracy the transit absorption in multiplet lines.

To demonstrate this method, we model atmospheres of selected hot exoplanets and calculate transit absorptions, comparing synthetic and available measured spectral features of multiplets with and without magnetic fields. We aim to describe the atomic alignment effect, applied for the first time for interpretation of transit absorptions in exoplanetary atmospheres. This theory is quite novel and rather complex in itself. To focus on this aim, we do not overload the paper with the detailed description of our numerical modeling, which can be found in cited references. The paper structure is the following. Key equations and theoretical base are presented in the Model Description section. The Results section discusses the differences between simulations with and without MFs and highlights promising multiplet lines. The summary of the study and a brief outline of the results are provided in the conclusion.

1. Model description
1.1 Atmospheric Modeling and Transit Absorptions

Numerical model of exoplanetary atmospheres solves the hydrodynamic equations for all atmospheric components on a three-dimensional global spherical grid, employing a self-consistent multi-fluid approach that includes aeronomy, plasma-photo-chemical reactions, cooling and heating mechanisms due to absorption of the parent star's radiation. The evolutionary development of the model has been detailed in a series of publications: gas-dynamic equations and the processes of planetary wind generation are discussed in Shaikhislamov+ (2016) and Dwivedi+ (2019). Three-dimensional modeling of stellar winds is presented in Khodachenko+ (2019), while methodologies for transit absorption calculations are outlined in Shaikhislamov+ (2018). Examples of modeling of transit absorption in various lines of such elements as H, He, C, O, Mg, Si can be found in Shaikhislamov+ (2020). Specific features of modeling the population and absorption of metastable helium are described in Shaikhislamov+ (2021) and subsequent works (e.g., Rumenskikh+ 2024).

After simulating the distributions of velocities, densities, and temperatures of the planetary atmosphere in the frame of the stellar system, the calculation of spectral absorption profiles is performed. This involves solving the radiative transfer equations, incorporating Doppler and Lorentz broadening caused by the motion and density of absorbing particles. The absorption cross-section is determined using the Voigt profile, which is approximated using the Tasitsiomi (2006) empirical approximation to optimize computational efficiency. To obtain the optical transit absorption signal, the model performs sequential integration along the line of sight, accounting for the column density of absorbing atoms (optical depth).

$$\tau(V) = \int_{L}^{-L} dz \cdot n_i \cdot \sigma_{abs}(V, V_z, T) \qquad (1)$$

The transit absorption profile is then calculated as:

$$\text{Absorption} = 1 - \frac{I_{transit,\nu}}{I_{out,\nu}} = \frac{1}{\pi R_{St}^2} \int (1 - e^{-\tau}) \cdot dS \qquad (2)$$

As evident from Equation (2), the calculation of absorption in a spectral line requires determining the velocity, density, and temperature of the absorbing material, which is performed for each particle species in the planetary atmosphere during the atmospheric structure modeling stage. The dependence on atmospheric parameters vanishes when the optical depth becomes much larger than unity. This consideration underlies the selection of most absorption lines presented in Table 1 — we primarily chose multiplets of light elements to ensure that the absorbing medium corresponds to the least dense layers of the atmosphere.

The resulting synthetic spectra are subsequently compared with observational data. The comparison between models and observations allows us to assess the accuracy of the model and to evaluate parameters of the atmosphere. In this work, we present transit absorption spectra for several exoplanets for which the most plausible atmospheric parameters had been preliminarily estimated. These spectra are used to apply the alignment effects and compare modelled absorption with and without the influence of the magnetic field. At this stage, to demonstrate the alignment effect itself, we assume that the magnetic field affects only the statistical weights of transitions within the multiplet lines. Thus, we do not include MF into the hydrodynamic simulation. This is justified for sufficiently small MF below 0.1 G at the planet surface (Shaikhislamov+ 2020).

### 1.2 Atomic alignment

As is well known (see, for example, Varshalovich (1970) and the references cited there), the resonant interaction of directed unpolarized radiation with atoms under certain conditions leads to an anisotropic distribution of atoms over the projections of the total angular momentum of the interacting energy levels. It is characterized by an irreducible (with respect to the SU (2) group) second-rank tensor $\rho_{2q}$ (Omont 1977). This type of anisotropy is usually called alignment. In the case of exoplanetary atmospheres, the general cause of alignment is the naturally directed unpolarized radiation of the parent star. It is important to note that, as shown below, alignment can reach significant values (~0.2) and play a crucial role even in the case when the radiation spectrum of the parent star in the vicinity of the resonance line is "flat", i.e. the width of the pumping spectrum is much larger than not only the width of an individual line, but also the fine splitting of the lines in the multiplet. In this case, the counteracting processes that reduce the magnitude of alignment are depolarizing collisions and decay (both radiative and collisional) of the bottom level of transition at which the alignment occurs. The difference in populations of fine sublevels of the bottom state due to different projections of the orbital momentum of electron in the equilibrium isotropic state and in the case of alignment is schematically shown in Fig.1. The arrow demonstrate how the populations change due to pumping by stellar radiation (solid arrows) and subsequent radiative decay (solid and dash arrows). Based on the analysis of physical processes, it is possible to formulate a set of necessary conditions for observational manifestations of the alignment effect:

- Anisotropic (directional) radiation: $\Theta/4\pi \ll 1$ (typical value for hot exoplanets ~ 0.005);
- The total angular momentum of the lower level $J_g \geq 1$;
- Rarefied medium. The characteristic collision frequencies $\nu$ are much lower than the rate of spontaneous decay $\gamma$ of the excited level:

  $\nu$ (~ $1$-$10^2$ Hz, for hot exoplanets) $\ll \gamma$ (~ $10^6 - 10^8$ Hz);

- The rate of optical pumping by the radiation of the star is much greater than the rate of decay of the anisotropy of the lower level $\Gamma$ (usually it is the life-time of the bottom level, either the collision frequencies $\Gamma \sim \nu$, or radiative decay). The dimensionless saturation parameter of the transition S (see formula (9) below) is proportional to the spectral density of the radiation flux of the star and by the order of magnitude varies within the range of $0.01 - 0.001$.

  $\gamma \cdot S > \Gamma$

The absorption of the stellar radiation by the planet atmosphere can be seen by observer at primary transits. At present, a lot of observational data of this kind has been obtained. In order to reduce the factors counteracting the effect of atomic alignment, we will focus on triplet lines that are formed by transitions from a long-lived states (metastable or ground) with a total angular momentum $J_g \geq 1$. We solve the quantum kinetic equations in the approximation of small saturation of the transition (S << 1), and the population of the upper level being much less than the population of the lower one, when it is possible to obtain a closed system for the density matrix of the lower level.

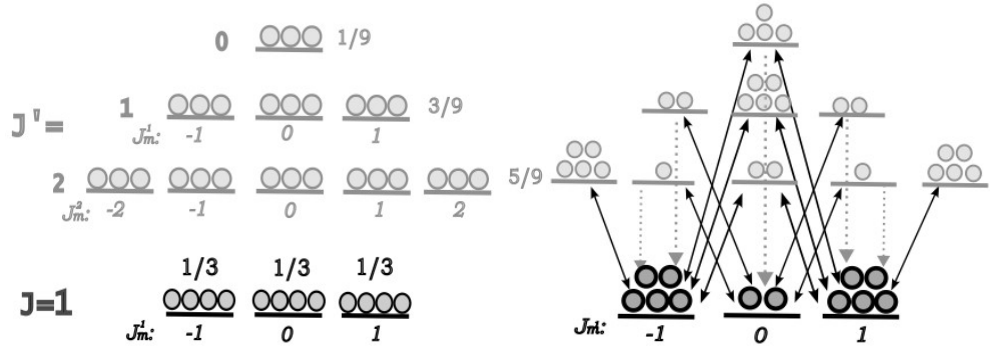

Figure 1. Schematic representation of a system of one lower and three upper sublevels with a fine structure of splitting by orbital momentum projection. The example of this scheme can be the transition $2^3S$ - $2^3P$ of HeI with the corresponding angular momenta Jg = 1 and Je = 0, 1, 2. On the left are the populations in the equilibrium state, when the probability obeys the relation 2J+1. On the right is the scheme with the populations due to interaction with the radiation of parent star.

## 1.3 Quantum mechanical formulation

Let's consider the resonant interaction of the directed unpolarized broadband radiation of a star with atoms whose energy levels have total angular momenta Jg in the lower long-lived state and Je in the upper short-lived states. Separating the fast dependence on time and coordinates, we write the electric field strength as

$$\mathbf{E}(t) = e^{-i\omega t + ikz}\mathbf{A}(t) + c.c.; \quad \mathbf{A}(t) = \sum_{q=\pm 1} A_q(t)\mathbf{e}_q; \quad \mathbf{e}_0 = \frac{\mathbf{k}}{k}; \quad \mathbf{e}_{\pm 1} = \pm \frac{-i\mathbf{e}_y - \mathbf{e}_x}{\sqrt{2}},$$

where the z- axis coincides with the direction of field propagation, the slow amplitudes $A_q(t)$ are stationary Gaussian processes with zero mean value $\langle A_q(t)\rangle = 0$. In this case, the spectral and polarization characteristics of the field are described by the (auto)correlation matrix $\Psi_{qq'}(t - t') = \langle A_q(t)A_{q'}^*(t')\rangle$. In particular, for an unpolarized field $\Psi_{qq'}(\tau) = \delta_{qq'}\Psi(\tau)$ and the Fourier transform of the function gives the spectral power density of the field (Wiener-Khinchin theorem).

In the non-relativistic approximation, the Hamiltonian of a free atom is the sum of the kinetic energy and the Hamiltonian of the atom in the system of its center of mass, which can be represented as an expansion in projection operators $\widehat{\Pi}_i = \sum_{mi} |J_i, m_i\rangle\langle J_i, m_i|$, where $|J_i, m_i\rangle$ is the state vector with given energy $E_i$, total angular momentum $J_i$ and its projection $m_i$ on the z axis :

$$\widehat{H}_A = \frac{\widehat{\mathbf{p}}^2}{2M} + \sum_{i=e,g} E_i \widehat{\Pi}_i$$

The Hamiltonian of the interaction of an atom with a radiation field in the dipole and resonance approximations has the form

$$\widehat{H}_{A-E} = e^{-i\omega t + ik\hat{z}} \sum_e d_{eg} \sum_{q=\pm 1} \widehat{D}_q(e) A_q(t) + h.c. \qquad (1)$$

where $d_{eg}$ is the reduced matrix element of the dipole moment of the transition Jg <→ Je, and the irreducible Wigner tensor operators of the first rank are defined by the formula

$$\widehat{D}_q(e) = \sum_{\mu,m} |J_e, \mu\rangle C^{J_e,\mu}_{J_g,m;1,q} \langle J_g, m| \qquad (2)$$

Here are $C^{J_e,\mu}_{J_g,m;1,q}$ the Clebsch-Gordan coefficients (Varshalovich+ 1975). In the presence of a static magnetic field, the operator **B** of the energy of interaction of the atom with the magnetic field is added to the Hamiltonians described above, which in the case of not very strong fields describes the linear Zeeman splitting of levels:

$$\widehat{H}_{A-B} = \sum_{i=e,g} \mu_B g_i (\hat{\mathbf{J}}_i \mathbf{B}) \qquad (3)$$

where $\mu_B$ is the Bohr magneton, $g_i$ the Lande gyromagnetic factor of level $i$, $\hat{\mathbf{J}}_i$ and is the operator of the total angular momentum of this level.

### 1.4 Quantum kinetic equation and reduction

The evolution of a single-particle density matrix $\hat{\rho}$ of an atom in a radiation field $\mathbf{E}(t)$ and a magnetic field **B** is described by the quantum kinetic equation:

$$\frac{\partial \hat{\rho}}{\partial t} = -\frac{i}{\hbar} [\widehat{H}, \hat{\rho}] + \widehat{\Gamma}\{\hat{\rho}\}$$

in which $\widehat{H} = \widehat{H}_A + \widehat{H}_{A-E} + \widehat{H}_{A-B}$ is the complete Hamiltonian of the atom, and the term (relaxation operator) $\widehat{\Gamma}\{\hat{\rho}\}$ describes the relaxation processes caused by both spontaneous emission of photons and collisions of various types. Usually, the concentration of the active gas participating in the formation of the absorption spectrum is small and the relaxation operator is linear with respect to $\hat{\rho}$. Its *form* can be derived within the framework of certain assumptions and approximations (see, for example, Rautian & Shalagin 1991), but in practice it is set from phenomenological considerations, which leads to the so-called relaxation constant model. In the present work we restrict ourselves to the approximation of purely radiative relaxation, when the form $\widehat{\Gamma}\{\hat{\rho}\}$ is well known (Taichenachev+ 2004).

The procedure for reducing the original quantum kinetic equation for the density matrix $\hat{\rho}$, which contains rapidly oscillating coefficients and random functions, to a closed system of kinetic equations for the level density matrices

$$\langle \hat{\rho}_{gg} \rangle = \hat{\sigma}_g; \quad \langle \hat{\rho}_{ee} \rangle = \hat{\sigma}_e$$

averaged over the ensemble realizations of a random field $\mathbf{E}(t)$ is described in sufficient detail in (Taichenachev+ 2004). Therefore, here we reproduce this system of equations without derivation, using the notation described above. Thus, for the density matrices of excited levels, we can write

$$\left[\frac{\partial}{\partial t}\right] \hat{\sigma}_e = -\frac{|d_{eg}|^2}{\hbar^2} \int d\omega \sum_{q,q'=\pm 1} \widetilde{\Psi}_{qq'}(\omega) \left\{ \frac{\widehat{D}_q(e)\widehat{D}^\dagger_{q'}(e)}{\frac{\gamma}{2}+i(\omega-\omega_{eg})} \hat{\sigma}_e + \hat{\sigma}_e \frac{\widehat{D}_q(e)\widehat{D}^\dagger_{q'}(e)}{\frac{\gamma}{2}-i(\omega-\omega_{eg})} - \frac{\gamma}{\left(\frac{\gamma}{2}\right)^2+(\omega-\omega_{eg})^2} \widehat{D}_q(e)\hat{\sigma}_g \widehat{D}^\dagger_{q'}(e) \right\} - \gamma \hat{\sigma}_e - i\Omega_e [(\hat{\mathbf{J}}_e \mathbf{b}), \hat{\sigma}_e] \qquad (4.1)$$

These equations must be supplemented by a quantum kinetic equation for the density matrix of the lower long-lived state

$$\left[\frac{\partial}{\partial t}\right]\hat{\sigma}_g = -\frac{|d_{eg}|^2}{\hbar^2}\int d\omega \sum_{e,q,q'=\pm 1}\tilde{\Psi}_{qq'}(\omega)\left\{\frac{\hat{D}_{q'}^\dagger(e)\hat{D}_q(e)}{\frac{\gamma}{2}-i(\omega-\omega_{eg})}\hat{\sigma}_g + \hat{\sigma}_g\frac{\hat{D}_{q'}^\dagger(e)\hat{D}_q(e)}{\frac{\gamma}{2}+i(\omega-\omega_{eg})}\right.$$

$$\left.-\frac{\gamma}{\left(\frac{\gamma}{2}\right)^2+(\omega-\omega_{eg})^2}\hat{D}_{q'}^\dagger(e)\hat{\sigma}_e\hat{D}_q(e)\right\} + \gamma\sum_{e,s=\pm 1,0}\hat{D}_s^\dagger(e)\hat{\sigma}_e\,\hat{D}_s(e) - i\Omega_{\rm g}[(\hat{J}_g\mathbf{b}),\hat{\sigma}_g] \qquad (4.2)$$

The terms in curly brackets describe the processes of optical pumping of levels and their light shifts in the field of broadband radiation specified by the (auto)correlation matrix

$$\Psi_{qq'}(t-t') = \int d\Omega e^{-i\Omega(t-t')}\tilde{\Psi}_{qq'}(\Omega) \qquad (5)$$

The operator of radiative relaxation of levels has a standard form and contains terms of "departure" $(-\gamma\hat{\sigma}_e)$ and "arrival" $(+\gamma\sum_{e,s=\pm 1,0}\hat{D}_s^\dagger(e)\hat{\sigma}_e\,\hat{D}_s(e))$, preserving the total population (closed system of levels). The last terms in (4.1) and (4.2) describe the Zeeman splitting of levels in a static magnetic field with the corresponding splitting frequencies:

$$\Omega_{e,g} = \frac{\mu_B g_{e,g} B}{\hbar}; \quad \mathbf{b} = \frac{\mathbf{B}}{B}$$

the Zeeman splitting of lines is neglected in view of the large characteristic width of the radiation spectrum $\Delta\omega \gg \Omega_i, k\bar{p}/M$.

### 1.5 Low saturation limit of the transition

As a first step, we consider the interaction of an ensemble of atoms with the radiation of a star without taking into account the magnetic field. In a weak radiation field, the (quasi)stationary distribution over the sublevels of the excited state adiabatically tracks the distribution in the lower metastable state:

$$\hat{\sigma}_e = \frac{|d_{eg}|^2}{\hbar^2}\int d\omega \sum_{q,q'=\pm 1}\frac{\tilde{\Psi}_{qq'}(\omega)}{\left(\frac{\gamma}{2}\right)^2+(\omega-\omega_{eg})^2}\hat{D}_q(e)\hat{\sigma}_g\hat{D}_{q'}^\dagger(e) \qquad (6)$$

In the same limit of small transition saturation, we obtain a closed equation for the density matrix of the ground state

$$\left[\frac{\partial}{\partial t}\right]\hat{\sigma}_g = -\frac{|d_{eg}|^2}{\hbar^2}\int d\omega \sum_{e,q,q'=\pm 1}\tilde{\Psi}_{qq'}(\omega)\left\{\frac{\hat{D}_{q'}^\dagger(e)\hat{D}_q(e)}{\frac{\gamma}{2}-i(\omega-\omega_{eg})}\hat{\sigma}_g + \hat{\sigma}_g\frac{\hat{D}_{q'}^\dagger(e)\hat{D}_q(e)}{\frac{\gamma}{2}+i(\omega-\omega_{eg})}\right.$$

$$\left.-\frac{\gamma}{\left(\frac{\gamma}{2}\right)^2+(\omega-\omega_{eg})^2}\sum_{s=\pm 1,0}\hat{D}_s^\dagger(e)\hat{D}_q(e)\hat{\sigma}_g\hat{D}_{q'}^\dagger(e)\,\hat{D}_s(e)\right\} \qquad (7)$$

with the normalization condition $\text{Tr}\{\hat{\sigma}_g\} = const = 1$. For unpolarized radiation $\tilde{\Psi}_{qq'}(\omega) = \delta_{qq'}\tilde{\Psi}(\omega)$, the equation for the ground state density matrix is reduced to the form:

$$\left[\frac{\partial}{\partial t}\right]\hat{\sigma}_g = \sum_{e,q=\pm 1}\gamma S_e\left(-\frac{1}{2}\{\hat{D}_q^\dagger(e)\hat{D}_q(e),\hat{\sigma}_g\} + \sum_{s=\pm 1,0}\hat{D}_s^\dagger(e)\hat{D}_q(e)\hat{\sigma}_g\hat{D}_q^\dagger(e)\,\hat{D}_s(e)\right) \qquad (8)$$

where we introduced into consideration the dimensionless partial saturation parameter of the transition Jg → Je

$$S_e = \frac{|d_{eg}|^2}{\hbar^2} \int d\omega \frac{\Psi(\omega)}{\left(\frac{\gamma}{2}\right)^2 + (\omega - \omega_{eg})^2} \tag{9}$$

If the width of the emission spectrum significantly exceeds the value of the fine splitting of the multiplet, then the saturation parameter is the same for all Je. It is important to note that the saturation parameter determined in this way does not depend on the value of the transition dipole moment and is determined in fact by the average number of photons in the emission mode at the transition frequency. For instance, an estimate of the saturation parameter for the transition from the metastable state of the He I at a wavelength of $\lambda = 1083$ nm for typical spectral flux density of $10^3$ (erg/cm $^2$ /s /A) yields $S_e \sim 10^{-3}$.

## 2. Results

### 2.1 Analytical results

For small values of the total angular momentum Jg, the solution of the differential system of equations (4.1) and (4.2) can be found in explicit analytical form for any initial state. However, the analytical expressions become simpler and the physical meaning of the solutions obtained is clearer if we use the following technique. We replace differentiation with respect to time in (7) with a finite difference $\left[\frac{\partial}{\partial t}\right]\hat{\sigma}_g \to \frac{\hat{\sigma}_g - \hat{\sigma}_g(0)}{\tau}$, where $\hat{\sigma}_g(0) = \hat{\Pi}_g/(2J_g + 1)$ is the isotropic distribution over the sublevels of the ground state in the absence of a radiation field, and $\tau = 1/\Gamma$ has the meaning of the lifetime of the metastable state, which in our case is determined by collisions. After such replacement, the problem is reduced to solving a system of linear algebraic equations, which is easy to find for any values of the dimensionless parameter $\gamma S_e \tau$. For example, the alignment of the metastable state of the helium atom

$$Q = (\sigma_{-1} + \sigma_{+1} - 2\sigma_0)/\sqrt{6} = \frac{5\gamma S_e \tau}{\sqrt{6}(18 + 11 \quad_e \tau)} \tag{10}$$

where $\sigma_m$, m=0, ±1 are the populations of the Zeeman sublevels of the metastable state. As expected, the alignment is maximal under conditions where the optical pumping rate is much greater than the relaxation rate of the metastable state $\gamma S_e \tau \gg 1$ and reaches the limiting (stationary) value $Q_{st} = \frac{5}{11\sqrt{6}} = 0.19$. This alignment is manifested in the transit absorption spectra at a wavelength of $\lambda = 1083$ nm in the form of a deviation of the relative intensities of the absorption lines in the multiplet from the equilibrium values, which are proportional to the statistical weights of the corresponding excited levels $(2J_e + 1)$ (Sobel'man 1963). In particular, the relative intensity of the absorption line with $J_e = 0$ in our case will be determined by the expression

$$I(J_e = 0) = \frac{4 + 3\gamma S_e \tau}{36 + 22 \quad_e \tau} \tag{11}$$

from which it follows that with an increase in the parameter, $\gamma S_e \tau$ the relative absorption intensity in this component of the multiplet increases from 1/9 (equilibrium value) to 3/22 (stationary value). This difference is accessible to observation by modern spectrographs.

In the presence of an arbitrarily oriented magnetic field, equations (6) and (7) must be supplemented by the corresponding terms describing the linear Zeeman splitting of levels. Finding a solution in this case is quite simple being reduced to solving a system of linear algebraic equations. However, the resulting explicit analytical expressions are quite cumbersome and the results are best analyzed graphically. It is important to note that even in a relatively weak magnetic field of the order of 1 G, the Zeeman splitting frequency significantly exceeds the optical pumping rate of the metastable level $\Omega_g \gg \gamma S_e$. This leads to the fact that the alignment formed by unpolarized radiation in the presence of a magnetic field changes significantly both in magnitude and in its orientation (the symmetry axis will now be determined by the magnetic field, and not by the direction of radiation propagation).

## 2.2 Application of alignment theory

Detecting spectrally resolved transit absorptions of hot exoplanets is technically challenging. At present, detections are limited to lines with high oscillator strengths, arising from dipole-type transitions. In these transitions, the angular momentum can change by ±1 or remain constant. Consequently, we focus on multiplets where the electronic angular momentum quantum numbers satisfy the relations Je=Jg, Je=Jg±1 (e.g., Jg=1 → Je=0,1,2, Fig. 1, or Jg=5/2 → Je=3/2,5/2,7/2, etc). These relations correspond to different types of transitions, and the alignment effect acts differently on the level populations associated with these transitions.

Despite fulfilling the conditions for observing the alignment effect (as described in Section 3), several specific challenges arise in the context of exoplanet transit absorption lines. First, these lines are typically broad, with full widths at half maximum (FWHM) reaching several angstroms. This imposes a constraint on the triplet characteristics, requiring each line to be separated by at least 1 Å. Second, atmospheric stratification suggests that heavy elements are concentrated closer to the planet in dense atmospheric layers. Because the method works in optically thin media, only moderately heavy elements are suitable. Third, transitions between upper and lower levels must be faster than other transitions which can occur with these levels. Therefore, we focused on transitions originating from long-lived levels to identify suitable lines. Table 1 lists atomic lines satisfying the criteria for observing the alignment effect, along with their key characteristics. Provided the atmospheric optical depth in the absorption region remains low, these lines represent promising targets for observations to infer the magnetic fields of exoplanets.

| Atom or ion | Term and configuration of lower level | Term and configuration of upper level | Lower and upper level energy, eV | Jg | Transition wavelengths | Lifetime of the lower level, s | Statweight without alignment | Statweight with alignment |
|---|---|---|---|---|---|---|---|---|
| He I | 1s2s $^3$S | 1s2s $^3$P | 19.82 20.96 | 1 | 10829.9 10830.02 10830.03 | 8000 | J→J-1 0.111 J→J 0.333 J→J+1 0.556 | J→J-1 0.136 J→J 0.295 J→J+1 0.568 |
| Na I | 2p$^6$3d $^2$D | 2p$^6$9f $^2$F° | 3.62 4.97 | 5/2 | J→J-1 9153.8 J→J 9153.8 J→J+1 9412 | 0.3 | J→J-1 0.222 J→J 0.333 J→J+1 0.444 | J→J-1 0.177 J→J 0.299 J→J+1 0.492 |
| Si I | 3s$^2$3p$^2$ $^1$D | 3s$^2$3p3d $^3$D° | 0.78 6.73 | 2 | J→J-1 2087.6 J→J 2086.7 J→J+1 2084.5 | 7000 | J→J-1 0.2 J→J 0.333 J→J+1 0.467 | J→J-1 0.159 J→J 0.295 J→J+1 0.5 |
| S I | 3s$^2$3p$^4$ $^1$D | 3s$^2$3p$^3$(2D°)4s $^3$D° | 1.15-8.41 | 2 | J→J-1 1706.3 J→J+1 1707.1 | 6200 | J→J-1 0.375 J→J+1 0.625 | J→J-1 0.159 J→J+1 0.439 |
| Ti II | 3d$^2$(1G)4s $^2$G | 3d$^2$(1G)4p $^2$H° | 1.89-5.69 | 9/2 | J→J 3286 J→J+1 3261 | 300 | J→J 0.33 J→J+1 0.4 | J→J 0.3 J→J+1 0.48 |
| Fe II | 3d$^5$4s$^2$ a $^6$S | 3d$^6$(5D)4p z $^6$P° | 2.89 5.36 | 5/2 | J→J-1 4924 | ~0.6 | J→J-1 0.222 | J→J-1 0.177 |

| | | | | J→J 5018 J→J+1 5169 | | J→J 0.333 J→J+1 0.444 | J→J 0.299 J→J+1 0.492 |

*Table 1. Promising multiplets for observing the alignment effect*

As discussed in Sect. 3, the directed radiation flux from the host star aligns the lower-level populations. Thus, the anisotropic radiation field alters the relative absorption intensities of individual lines in the triplet, deviating from the 2J+1 statistical weight distribution. However, the presence of a magnetic field destroys this anisotropy by "rotating" the quantization axis. For magnetic fields $\geq 10^{-3}$ G, the fine structure level populations return to the equilibrium 2J+1 ratio (Fig. 2). This property can be utilized to diagnose the absence of magnetic fields, as the most significant deviations from the 2J+1 ratio occur under field-free conditions. For example, if the weakest line in NaI multiplet is measured to have relative absorption at a level of 0.177 and less than the equilibrium value of 0.222, it means that we indeed observe the alignment effect not affected by other factors such as the optical depth, collisions and the magnetic field.

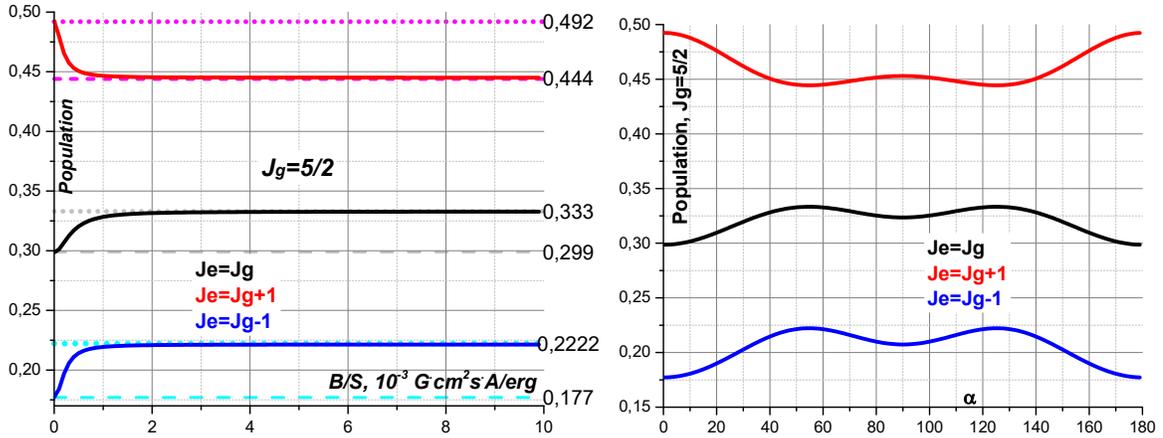

*Figure 2. Population distributions of upper levels for Jg=5/2 as a function of the magnetic field-to-radiation flux ratio (left) and the angle α between the quantization axis (star-planet axis) and the magnetic moment direction (right)*

It is important to note that the populations revert back to equilibrium only when the angle α between the quantization axis (star-planet axis) and the magnetic moment direction is non-zero. As α increases from 0° to approximately 50°, populations approach equilibrium, displaying a harmonic angular dependence. A secondary maximum occurs near α~140°. Importantly, the alignment effect depends not on the magnetic field's sign but on the specific angle α\alphaα. While the degeneracy between the magnetic field and radiation flux complicates parameter estimation, analyzing multiple absorption lines during a single transit can partially resolve this issue. Anyway, in practice, for dipole or quadrupole field, as most probable, whatever the orientation of magnetic moment the largest volume of atmosphere will be under magnetic field directed perpendicular to the planet-star line. Thus, in the first approximation, namely the value of magnetic field is important and could be distinguished in observations of alignment effect, regardless of its orientation.

To assert that the planetary magnetic field exceeds the value of 0.001·S when we observe near equilibrium populations require stricter constrains. We should be sure that the processes which can destroy the alignment effect do not operate – collisional and photo-induced transitions, large optical depth.

Generally, larger angular momentum values produce more significant deviations from the equilibrium 2J+1 distribution under alignment effect. It is more prominent for transitions involving higher angular momentum states, as the number of magnetic sublevels grows, enhancing the sensitivity to alignment. Figure 3 provides a detailed illustration of how the upper-level populations for triplet lines depend on the lower-level angular momentum Jg. Notably, the disparities between the Jg→Je+1 and Jg→Je−1 transitions become progressively larger as Jg increases, particularly in the absence of magnetic fields. These deviations result from the breaking of statistical equilibrium under directional irradiation, which preferentially populates certain magnetic sublevels, thereby altering the relative strengths of the individual components.

Proposed method not only highlights the role of alignment effects in reshaping absorption line profiles but also broadens the scope of alignment-sensitive spectral diagnostics. By examining lines with higher angular momentum values, where deviations are more pronounced, it becomes possible to identify new candidates for probing alignment processes in planetary atmospheres and other astrophysical environments. Such transitions offer additional tools for testing atmospheric models, as the degree of alignment sensitivity is tied to the interplay between anisotropic irradiation, local thermal conditions, and atmospheric dynamics.

Importantly, Figure 3 serves as a practical guide for expanding the current list of alignment-sensitive transitions. Readers may use this figure in conjunction with spectral databases, such as NIST, and the Table 1 to conduct independent searches for lines with suitable angular momentum characteristics. Lines featuring significant deviations from equilibrium populations are prime targets for future observational campaigns aimed at detecting and quantifying alignment effects. These transitions may reveal new insights into the upper-atmosphere physics of exoplanets, including the influence of stellar radiation fields, magnetic effects, and planetary outflows, all of which contribute to the observed line shapes and intensities. Note that a system with Jg=1, like HeI triplet, shows the opposite behavior for the component Jg→Je−1 with the lowest population: population is larger with alignment than the equilibrium value.

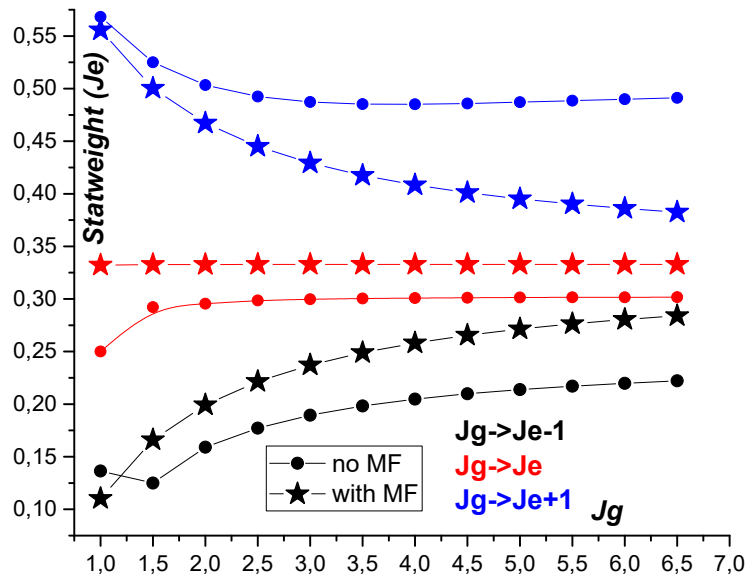

*Figure 3. Relative statweights of upper triplet sublevels of quantum system in the presence (asterisks) and absence (circles) of magnetic field as a function of the lower-level angular momentum Jg.*

Figure 4 presents the calculated absorption lines considering the absence and presence of intrinsic planetary magnetic fields for two planetary systems: HD 189733 and WASP-69. The assessment of the physical and chemical parameters of these systems was previously conducted and is described in detail in the works of Rumenskikh+ (2022) and Rumenskikh+ (2024). As can be observed, the helium triplet absorption in the WASP-69b system significantly differs from that of HD 189733b, with notably lower absorption in the weakest component of the triplet. In the frame of proposed theory, this can be explained by the planet's intrinsic magnetic field. However, to assert this, we need to verify that collisions are not important. Such verification should involve full simulation of atmosphere with assumed magnetic field.

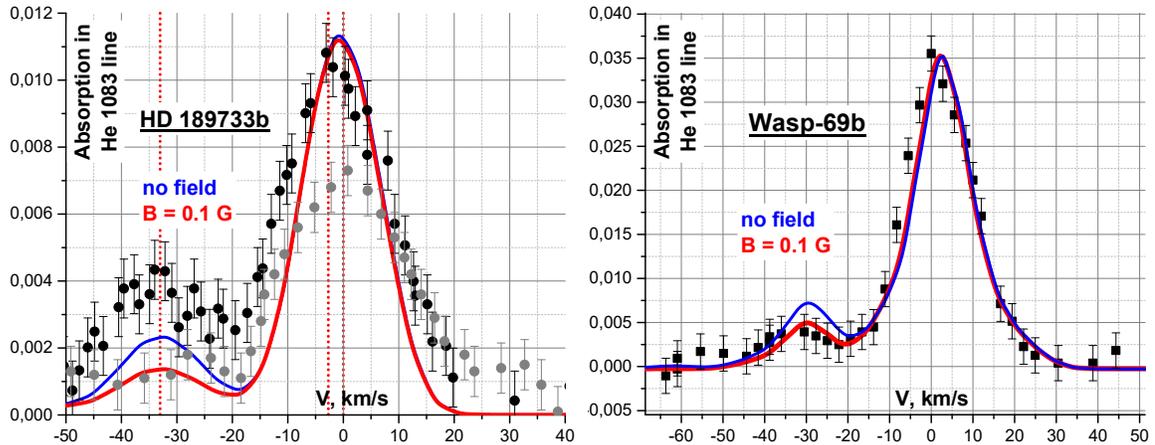

*Figure 4. Observed absorption in the metastable helium triplet for HD 189733 b and Wasp-69 b (dots), compared to model predictions with (red) and without (blue) magnetic fields. Wavelength is plotted in Doppler velocities.*

In the case of HD 189733 b, observations of the helium triplet lines exhibit significant variability, particularly in the J→J−1 transition line, which has been the subject of several independent studies. Notably, data reported by Salz+ (2019) appear to align with the predictions of a field-free case, where the absorption profiles can be explained solely by atmospheric dynamics, thermal effects, optical depth and radiative processes without the influence of a planetary magnetic field. In contrast, observations presented by Guilluy+ (2020) suggest the presence of a detectable magnetic field, as their measured line ratios deviate from the field-free predictions. The discrepancies between these studies highlight the complex nature of the system and the sensitivity of the helium triplet lines to magnetic and environmental conditions, underscoring the challenges associated with disentangling the contributing factors.

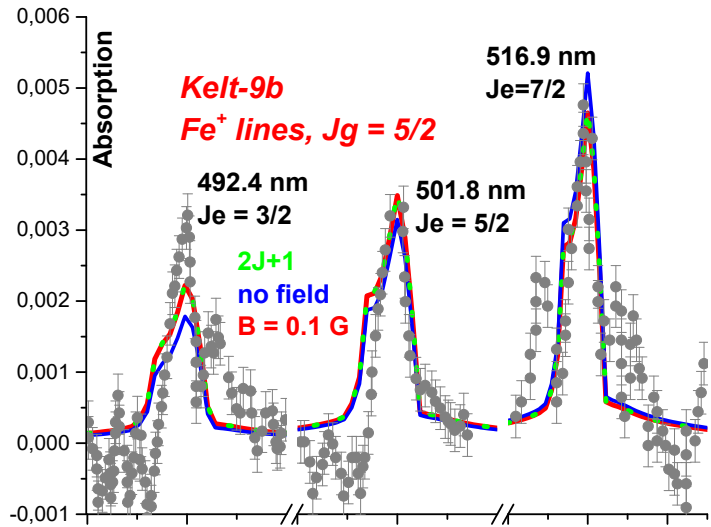

*Figure 7. Observed (dots) and modeled (lines) absorption in iron triplet lines for Kelt-9 b, with (red) and without (blue) magnetic field of the planet*

Aanalysis of other triplet lines also show interesting results. For instance, transit observations of Kelt-9 b demonstrate significant absorption in iron triplet at around 500 nm, which coincides well with the theoretical predictions summarized in Table 1. When intrinsic magnetic field of the planet is taken into account in calculating the absorption, the resulting profiles show a notably improved agreement with the measurement. However, the match is not perfect, as some deviations remain, particularly in peak intensities. These discrepancies are likely attributable to the limited signal-to-noise ratio and uncertainties in the observational data. Nevertheless, within the bounds defined by measurement noise and observational precision, the agreement can be considered satisfactory, lending further credibility to the inclusion of planetary magnetic fields in the models.

**Conclusions**

The theory of atomic alignment, known in other areas of optics, is applied for the first time to interpret transit absorptions of hot exoplanets. We shew that it's possible to trace the presence or absence of intrinsic magnetic fields by accurately measuring the absorption in individual components in triplet lines. Due to directed stellar radiation a redistribution of atoms in sublevels with different projections of orbital momentum of the lower level occurs. If the lower level is sufficiently long-lived, than multiple acts of radiative excitation and de-excitation between lower and upper levels results in unequal populations of sublevels of lower level differentiated by projections of orbital momentum. This results in significantly different statweights of absorption probability in individual components of multiplet transition between lower and upper levels, than the equilibrium ratio 2J+1. However, the presence of magnetic field disrupts this effect: for sufficiently large values of the magnetic field not in parallel with planet-star line, the anisotropy of the radiation field inducing alignment is disrupted by the rotation of the quantization axis, and the statweights of absorption probability again return to the equilibrium ratio 2J+1. Thus, the method presented in this paper can serve, first of all, to determine the absence of magnetic field of exoplanets than otherwise. However, with improved observation quality and detailed modeling, the presence of even rather small magnetic field can be detected.

We presented analyses of some available typical measurements of absorption by suitable triplets in the context of the alignment effect. While in general they do not contradict the alignment theory and even support some definite conclusions on absence or presence of magnetic field, one can see that the quality of observations should improve by several times before conclusive results might be obtained. Besides S/N ratio, future observations require long-term monitoring in view of variability of stellar and atmospheric parameters. Observations of more than one triplet at the same time can eliminate some crucial degeneracy, helping in detecting planetary magnetic field.

**Data availability**

The data underlying this article will be shared on reasonable request to the corresponding author.

**Acknowledgements**

This work has benefited from the contributions of many individuals, and I would like to express my gratitude to everyone who showed interest and provided valuable advice during the idea's formation and its presentation at scientific conferences. Special thanks go to Roman Kislov, Mark Blumenau, and Mikhail Fridman, whose discussions inspired the idea of searching for exoplanetary magnetic fields through the analysis of absorption lines. I am also deeply grateful to O. N. Prudnikov for assistance in elucidating the nuances of numerical modeling for systems of quantum-kinetic equations. Finally, we extend our sincere appreciation to the Russian Science Foundation for supporting this research under grant № 23-12-00134.

**References**

Bourrier , V., Des Etangs , A. L., & Vidal- Madjar , A. (2014). Modeling magnesium escape from HD 209458b atmosphere. *Astronomy & Astrophysics*, *565*, A105.

Oklopčić , A., & Hirata, C. M. (2018). A new window into escaping exoplanet atmospheres: 10830 Å line of helium. *The Astrophysical Journal Letters* , *855* (1), L11.

Lampon M.+ Modeling the He I triplet absorption at 10,830 Å in the atmosphere of HD 209458 b //Astronomy & Astrophysics. – 2020. – T. 636. – P. A13.

Odert , P., Erkaev , N.V., Kislyakova , K.G., Lammer, H., Mezentsev , A.V., Ivanov, V.A., ... & Holmström , M. (2020). Modeling the Ly α transit absorption of the hot Jupiter HD 189733b. *Astronomy & Astrophysics* , *638* , A49.

Wang, L., & Dai, F. (2021). Metastable Helium Absorptions with 3D Hydrodynamics and Self-consistent Photochemistry. II. WASP-107b, Stellar Wind, Radiation Pressure, and Shear Instability. *The Astrophysical Journal* , *914* (2), 99.

Ben-Jaffel , L., Ballester , G. E., Muñoz, A. G., Lavvas , P., Sing, D. K., Sanz- Forcada , J., ... & López-Morales, M. (2022). Signatures of strong magnetization and a metal-poor atmosphere for a Neptune-sized exoplanet. *Nature Astronomy* , *6* (1), 141-153.

Khodachenko , M. L., Shaikhislamov , I. F., Lammer, H., Miroshnichenko , I. B., Rumenskikh , M. S. , Berezutsky , A. G., & Fossati , L. (2021). The impact of intrinsic magnetic field on the absorption signatures of elements probing the upper atmosphere of HD209458b. *Monthly Notices of the Royal Astronomical Society* , *507* (3), 3626-3637.

Lammer, H., Selsis , F., Ribas , I., Guinan, E. F., Bauer, S. J., & Weiss, W. W. (2003). Atmospheric loss of exoplanets resulting from stellar X-ray and extreme-ultraviolet heating. *The Astrophysical Journal* , *598* (2), L121.

Murray-Clay, R. A., Chiang, E. I., & Murray, N. (2009). Atmospheric escape from hot Jupiters. *The Astrophysical Journal*, *693* (1), 23.

Zarka, P. (2007). Plasma interactions of exoplanets with their parent star and associated radio emissions. *Planetary and Space Science*, *55* (5), 598-617.

Omont, A. (1977). Irreducible components of the density matrix. Application to optical pumping. Progress in quantum electronics, 5, 69-138.

Lynch, C. R., Murphy, T., Lenc, E., & Kaplan, D. L. (2018). The detectability of radio emission from exoplanets. *Monthly Notices of the Royal Astronomical Society*, *478* (2), 1763-1775.

Turner, J. D., Zarka, P., Grießmeier, J. M., Lazio, J., Cecconi, B., Enriquez, J. E., ... & de Pater, I. (2021). The search for radio emission from the exoplanetary systems 55 Cancri, υ Andromedae, and τ Boötis using LOFAR beam-formed observations. *Astronomy&Astrophysics*, *645*, A59.

Varshalovich, D. A. (1970). Spin state of atoms and molecules in the cosmic environment. *Uspekhi fizicheskikh nauk*, *101* (7), 369-383.

Yan, H., & Lazarian, A. (2008). Atomic alignment and diagnostics of magnetic fields in diffuse media. *The Astrophysical Journal*, *677* (2), 1401.

Taichenachev, A. V., Tumaikin, A. M., Yudin, VI, & Nienhuis, G. (2004). Steady state of atoms in a resonant field with elliptical polarization. Physical Review A—Atomic, Molecular, and Optical Physics, 69(3), 033410.

Rumenskikh, MS, Shaikhislamov, IF, Khodachenko, ML, Lammer, H., Miroshnichenko, IB, Berezutsky, AG, & Fossati, L. (2022). Global 3D Simulation of the Upper Atmosphere of HD189733b and Absorption in Metastable He i and Lyα Lines. The Astrophysical Journal, 927(2), 238.

Rautian and A. M. Shalagin, *Kinetic Problems of Nonlinear Spectroscopy*